\documentclass[onecolumn,usenatbib]{Indulekha}
\usepackage{amsmath}
\begin{document}
\title{The role of gas dynamical friction in the evolution
of embedded stellar clusters} 
\author{K. Indulekha}
\institute{\center{School of Pure and Applied Physics, Mahatma Gandhi University,
Kottayam 686560, Kerala, India\\\
\email{E-mail: kindulekha@gmail.com, 
indulekha@mgu.ac.in}}}
\abstract
{Two puzzles associated with open clusters have attracted a lot of attention -- their formation, with densities and velocity dispersions that are not 
too different from those of the star forming regions in the Galaxy, given that the observed Star Formation Efficiencies (SFE) are low and, the mass 
segregation observed / inferred in some of them, at ages significantly less than the dynamical relaxation times in them. Gas dynamical friction has 
been considered before as a mechanism for contracting embedded stellar clusters, by dissipating their energy.  This would locally raise the SFE which 
might then allow bound clusters to form.  Noticing that dynamical friction is inherently capable of producing mass segregation, since here, the 
dissipation rate is proportional to the mass of the body experiencing the force, we explore further, some of the details and implications of such a 
scenario, vis-a-vis observations.  Making analytical approximations, we obtain a boundary value for the density of a star forming clump of given mass, 
such that, stellar clusters born in clumps which have densities higher than this, could emerge bound after gas loss.  For a clump of given mass and 
density, we find a critical mass such that, sub-condensations with larger masses than this could suffer significant segregation within the clump. }
\maketitle
\section{Introduction}
Stellar clusters can be viewed out to far distances and hence are important probes for our study of the universe. Most stars seem to form in clusters, and 
hence, understanding clusters becomes an essential requirement for understanding the formation and evolution of galaxies. Two puzzles associated with open 
clusters that have attracted a lot of attention are -- their formation, with densities and velocity dispersions that are not too different from those of the 
star forming regions in the Galaxy, given that the observed Star Formation Efficiencies (SFE) are low \citep{Tutukov1978, Hills1980, Mathieu1983, 
Elmegreen1983, LadaMargulisDearborn1984, Adams2000, GeyerBurkert2001, BK2007, Goodwin2009} and, the mass segregation observed / inferred in some of 
them \citep{ deGrijs2002, Sirianni2002, Gouliermis2004, Chen2007, ConverseStahler2010, HasanHasan2011}, at ages significantly less than their dynamical 
relaxation times \citep{BD1998, McMillanetal2007, Allisonetal2009, Allison2010, MASCHBERGER2011, OLCZAKSPURZEMHENNING2011}. In this paper, by making 
analytical approximations, we explore the efficacy of gas dynamical friction, operating during the  embedded phase of stellar clusters, in solving these 
two puzzles.\\\
Star clusters form, hidden from sight, embedded in the dense, dark and cold cores of giant molecular clouds \citep{LadaLada2003}. A star forming gas 
cloud converts less than ten percent of its mass into stars and then, in less than ten million years \citep{PS2000} loses the gas which had been 
gravitationally binding the stellar cluster.  A cluster of stars will be bound if the total energy of the stars is negative \citep{ChandrasekharElbert1972}. For mass loss that is instantaneous, analytical approximations \citep{Hills1980} show that the Star Formation Efficiency (SFE), i.e. the fraction of the total mass of the gas cloud that is converted into stars, should be greater than $0.5$, if a bound cluster is to emerge after gas loss. On the other hand, if the mass loss is gradual, bound clusters similar to those observed could emerge for SFE's greater than $\sim0.3$ \citep{Mathieu1983, Elmegreen1983, LadaMargulisDearborn1984}. Note however that this value is still much higher than the observed values for the global SFE. The SFE in dense cores which are forming clusters seem to be $\sim0.2 -0.3$, while the observed values for the SFE in star forming clouds are an order of magnitude smaller \citep{LadaLada2003}.\\\
Observations show also that, embedded clusters can start showing mass segregation in less than a few million years \citep{SchmejaKumarFerreira2008} and that the Orion Nebula Cluster is not old enough for the observed segregation of its massive stars to be due to dynamical relaxation by two body encounters alone \citep{BD1998}. Simulations indicate that the Pleiades cluster started out with significant mass segregation in the embedded phase itself, with high mass stars preferentially concentrated towards the center \citep{ConverseStahler2010}. The dynamical relaxation time in Pleiades is twice the present age of the cluster. Also, at the same time, some clusters like Taurus and Trumpler 16 do not show mass segregation \citep{Parkeretal2011, Wolketal2011, HasanHasan2011}.\\\
Numerical investigations of the dependence of the survivability of a cluster on the Star Formation Efficiency has shown that, a significant fraction of the stars can remain bound, after rapid gas loss, for an SFE $\gtrsim0.3$ and, for slow gas removal, up to $50\%$ of the cluster can remain bound for SFE values down to $0.15 -0.2$ \citep{GeyerBurkert2001, BK2007, Goodwin2009}.  However, if most open clusters are remnants of embedded clusters that have -- consequent to gas removal -- lost one half or more of their natal members \citep{BASTIAN2006, GOODWIN2006}, the similarity in the mass functions of embedded and open clusters and, the similarity in the IMF's of embedded and open clusters become difficult to understand \citep{LadaLada2003}.  In this context, scenarios in which bound clusters can form, without losing too much of their natal stellar content and, with (possibly) mass segregation become interesting.\\\
Gas dynamical friction due to an embedding gaseous medium has been invoked, in understanding many phenomena; at various scales -- from that of planetary systems to that of clusters of galaxies \citep{KimKim2009}. In the case of stars,  while considering mechanisms that change the momenta of stars in stellar systems, \citet{Chumak1976} conclude that, dynamical friction in a dust cloud would be the dominant momentum changing mechanism for a star, when there are no massive scattering centres and the speed of the star does not exceed a limit. Making use of the Chandrasekhar
dynamical friction formula \citep{Chandrasekhar1943, Chandrasekhar1943{a}}, \citet{SaiyadpourDeissKegel1997} had explored  gas dynamical friction as a means to produce contraction of embedded clusters, thus raising the SFE locally, which could then lead to the formaion of bound clusters. They had suggested also that, the same might be responsible for mass segregation \citep{Deiss1998}. \citet{GortiBhatt1995} had explored the effect of gas dynamical friction on pre-stellar clumps by numerical simulations and also, it has been noticed that gas dynamical friction can reduce the time scale for dynamical mass segregation \citep{Eretal2009}.  However, no general analysis of such a scenario, or its details and implications vis-a-vis observations, has been made.  Here, using analytical approximations, we explore a scenario in which gas dynamical friction could be playing a significant role, in the early evolution of embedded stellar clusters, as the chief mechanism which can cause them to contract and form more strongly bound configurations and, also cause mass segregation in them.  We check the plausibility of the model by examining its consistency with observations.\\\\
In section 2, using simple virial considerations, we check that, a scenario in which embedded clusters undergo contraction, meets the dynamical constraints set by observations.  In section 3 we consider dynamical friction in star forming gas clouds.  By requiring that the gas dynamical friction time scale within an embedded cluster be less than the time that is available before gas is expelled from the cluster, we obtain a boundary value for the density of a star forming clump of given mass, such that, stellar clusters born in clumps which have densities higher than this, could emerge bound after gas loss, and for a clump of given mass and density, we find a critical mass such that, subcondensations with larger masses than this could suffer significant segregation within the clump.  In section 4 we compare our results with observations. In section 5 we discuss some of the possible implications of our scenario and,  their accord with observations. In section 6 we give a brief summary of our results.  
\section{The SFE as a constraint on cluster formation scenarios -analytical approximations}
We now proceed as follows, for checking, whether the scenario we propose, wherein a nascent cluster contracts within the parent gas cloud, meets the 
constraints set by observations.
For a gas cloud that converts a fraction $\epsilon$ of its mass into stars and then loses the remaining gas,  \citet{Hills1980} had, by simple virial 
considerations, obtained expressions for the ratios of, the radius ($r_0$), velocity dispersion ($\sigma_0$) and density ($\rho_0$) of the gas cloud, 
with the radius ($r_f$), velocity dispersion ($\sigma_f$) and density ($\rho_f$)  of the stellar cluster, assumed virialized after gas loss (see also 
\citep{Mathieu1983}), as $\sigma_f / \sigma_0 = \sqrt \epsilon / \sqrt{r_f / r_0}$ and $\rho_f / \rho_0 = \epsilon / ({r_f / r_0})^3$, where $r_f / r_0$ 
equals $\epsilon /(2\epsilon -1)$ for instantaneous gas loss and equals $1 / \epsilon$ for adiabatic gas loss.  Making a similar analysis for an 
embedded 
stellar cluster, with $\epsilon <0.5$, that contracts to a 
radius $r' = pr_0$, where $p$ the contraction factor is $< 1$ we get, $r_f / r_0 = (p / (1 - ((1-\epsilon)/\epsilon) p^3))$ for instantaneous gas 
loss 
and $r_f / r_0 = p(1 + ((1-\epsilon)/\epsilon) p^3)$, if the gas is lost adiabatically.\\\\
For the first two cases, we see from the corresponding expressions for $r_f / r_0$ that, in the case of instantaneous mass loss, bound clusters may form only 
for SFE's $>0.5$ and in the case of slow mass loss, bound clusters may form, technically, for all values of $\epsilon >0$. In the case where the embedded 
cluster contracts to within a radius $r'=pr_0$, we see that, even if gas loss is instantaneous, bound clusters may form, for all values of $\epsilon < 0.5$ 
also, if $p$ is $ < {((\epsilon / (1 - \epsilon))}^{1/3}$. For $p \leq {((\epsilon / (1 - \epsilon))}^{1/3}$, the contracted cluster will have an {\it 
apparent} SFE $\geq 0.5$  \citep{SaiyadpourDeissKegel1997}. \\\\
In a real situation, a cluster will be situated in a galaxy, and hence, will be able to retain 
only the stars that are within the tidal radius. Our analysis, thus may be applied to those clusters, that do not fill their tidal radii and also, are young 
enough for their dynamics to have not been affected by either internal processes, like mass loss from the stars or external processes like interactions with 
molecular clouds.  The typical value for the density of a star 
forming region is $\sim10^4\,\rm{cm^{-3}}$ \citep{Elmegreen1983, BerginTafalla2007}.  And, taking for typical clusters, a density enhancement of $50$ with 
respect to the density of field stars -- which is of the order of $4\times10^{-24}\,\rm{gcm^{-3}}$ \citep{Allen1973} -- we get a typical value 
for $\rho_f /\rho_0 \sim 0.005$. The typical value for the observed velocity dispersion for clusters is $\sim 0.5\,\rm{kms^{-1}}$ and for molecular clouds 
it is of the order of a few kilometer per second, giving a typical value for $\sigma_f / \sigma_0 \sim 1/4 -1/6$. \\\\
We now estimate a typical value for the contraction 
factor $p$, by obtaining an estimate for $r_f / r_0$ using observations as, $r_f/r_0 = 1.7 (\epsilon/{0.025})^{1/3} ((\rho_f/\rho_0)/{0.005})^{-1/3}$ and then, solv{\bf ing} for $p$ using the relation  $r_f / r_0  = (p / (1 - ((1-\epsilon)/\epsilon) p^3))$ (in the case of instantaneous gas loss) or,  $r_f / r_0  =  p(1+((1-\epsilon)/\epsilon)p^3))$ (in the case of adiabatic gas loss).  The estimate for $p$ may be used to calculate  $\epsilon _a = 1 / (1 + ((1 - \epsilon)/\epsilon)p^3)$, the {\it apparent} SFE of the cluster,  when it has contracted to such an extent that, if it loses the gas now, it can remain bound, with a radius $r_f$ (after virialization), such that $r_f/r_0 = 1.7$.  For $r_f/r_0 = 1.7$ and $\epsilon = 0.025$, we get the required values for the contraction factor $p$ as, $p = 0.278$ in the case of fast gas loss and $p = 0.426$ in the case of slow gas loss.  These values for $p$, correspond respectively to {\it apparent} SFE's $\sim 0.544$ and $0.25$ for the embedded stellar cluster, in its most contracted phase, just prior to losing the gas. \\\\
In our scenario, we would expect at least some embedded stellar clusters, to show such high ({\it apparent}) SFE's, when near to the end of their embedded phase.  
Stellar densities that imply SFE's upto $0.47$ have been reported, in the dense embedded clusters observed in the $\rho$ Ophiucus, IC348 and Orion-Trapezium star forming regions (\citet{WilkingLada1983} and references therein).  We interpret these, as apparent SFE's, arising from the contraction of the particular nascent cluster. Thus these high SFE's that have been observed, lend support to our scenario. \\\\
As $\epsilon$ is varied from $0.008$ to $0.08$ the estimates for the appropriate {\it apparent} SFE's, as given by our model, go from $0.54$ to $0.576$ for fast gas loss and, from $\sim 0.2$ to $0.3$ for slow gas loss.  Assuming that $r_f/r_0$ has a larger value,  will make the {\it apparent} SFE values that correspond to it smaller, and vice versa.  As an aside we notice that, the value for the ratio $\sigma_f / \sigma_0$, calculated from the observed value for the ratio of the densities 
as, $\sigma_f / \sigma_0 =  1/8(\epsilon/{0.025})^{1/3}((\rho_f/\rho_0)/{0.005})^{1/6}$ is consistent with its value determined independently from direct observations -- a reflection of 
the fact that molecular clouds, as well as bound clusters, are both roughly in  virial equilibrium, a condition which was assumed in deriving these relations.\\\
Thus, we see that, our results linking $r_f/r_0, \sigma_f / \sigma_0$ and $\rho_f/\rho_0$, obtained under the assumption that the stellar cluster contracts within the parent gas cloud, are broadly consistent with observations. This conclusion of ours, is not very sensitive to the precise value of $\epsilon$ used in the calculations, as long as it is low.  With the above conclusion as our motivation, we now explore, by making analytical approximations, the efficacy of gas dynamical friction, in bringing about such a contraction of embedded stellar clusters. 
\section{The gas dynamical friction time-scale}
Molecular clouds which are observed to be the sites of star and 
cluster formation, have a hierarchical structure, of dense cores within clumps, the clumps themselves being within clouds -- the density getting lower 
as the scale becomes larger, from cores to clumps to clouds (see for example \citep{BerginTafalla2007}). For a clump of mass $m$, and velocity $\bf v$, moving in a medium with density $\rho_{gas}$, the gas dynamical friction time scale $t_{gdf}$ 
is $\sim (E/(dE/dt))$, where $dE/dt $ is the rate of energy dissipation due to gas dynamical friction and $E$ is the kinetic energy of the clump.\\\\
The retarding force, when a massive perturber that is interacting gravitationally moves through a fluid, may be determined by considering the interaction 
of the body with its own gravitationally induced over-density
wake.  For density wakes with small amplitudes, \citet{Ostriker1999} using time-dependent linear perturbation theory, determined the drag force on a 
point mass perturber, moving in a straight line, through a uniform, infinite, gaseous medium. For a mass $m$, moving with a speed $v$, in a medium with 
density $\rho_{gas}$ and sound speed $c_s$, she obtained an expression for the force as 
\begin{equation}
{\bf{F}}_{Lin} = -\frac{4\pi\rho_{gas}(Gm)^2}{v^3} {\bf{v}} f(\rm{M})
\end{equation}
where $f(\rm{M}) = 1/2ln((1+\rm{M})/(1-\rm{M}))-\rm{M}$ for $\rm{M} < 1$ and $ = 1/2
ln(1-(1/{\rm{M}}^2))+ln(vt/r_{min})$ for $\rm{M} > 1$.
Here $\rm{M} = v/c_s$ is the Mach number, $t$ is the time for which the perturber has been moving in the medium, 
and $r_{min}$ is a minimum radius introduced to avoid singularity in the force evaluation.  Later workers have relaxed 
the constraint on the perturber being point-like, to a perturber which has no boundary but is merging into the surrounding medium; that of a linear 
trajectory to considering circular orbits; and that of a uniform medium to one that is radially stratified (Kim \& Kim 2009 and references therein). 
These authors found that equation (1) is generally applicable with appropriate changes to the factor $(vt/r_{min})$. \\\\
Thus we get
\begin{equation}
\frac{dE}{dt} = k{\bf F}_{Lin}.{\bf v}=k \frac{4\pi\rho_{gas}(Gm)^2}{v}f(\rm{M}), 
\end{equation}
where k is a numerical factor that accounts for departures from the linear theory.  For a clump, moving within a cloud of total 
mass $M_c$ and radius $R$, with a speed $v$, parametrized by the virial speed for the cloud as $v = \beta v_{virial}$, putting 
$\rho_{gas} =\epsilon_{gas}\rho$ where, the gas fraction $\epsilon_{gas} = M_{gas}/M_c$, $M_{gas}$ being the mass in the interclump medium 
and $\rho$ is $M_{c}/(4\pi/3)R^3$, we get  
\begin{equation}
t_{gdf} = \frac{\beta ^3v_{virial}^3R^3}{6kG^2m\epsilon _{gas}M_cf(\rm{M})}. 
\end{equation}
Real cores, clumps and clouds are not spherically symmetric, have density gradients, and also substructure at a variety of scales.  However, potentials are rounder and smoother than the underlying density distribution \citep{BK2007}, and relaxing the assumption of sphericity, or homogeneity, introduces only changes 
by factors of order unity in the various terms of the virial equation \citep{ChandrasekharElbert1972, SomKochhar1985, Verschueren1990}.  Hence, for the term $v_{virial}$, we use the expression for the virial speed for a smooth homogeneous sphere of mass $M_c$ and radius $R$ viz. $\sqrt{(3/5)(GM_c/R)}$, and write
\begin{equation}
t_{gdf} = \frac{1}{20}\sqrt{\frac{9}{5\pi}}\frac{\beta ^3(M_c/m)}{k\epsilon_{gas}f(M)} t_{cross} = 0.038\frac{\beta ^3(M_c/m)}{k\epsilon_{gas}f(\rm{M})} t_{cross},
\end{equation}
where $t_{cross} = (G\rho)^{-1/2}$ is the crossing time in the cloud. \\\\
We rewrite the expression for the gas dynamical friction time-scale (Eqn.4) as  
\begin{equation}
t_{gdf} = \tau \mu t_{cross}
\end{equation}
 where
\begin{equation}
\tau = \frac{0.038 \beta ^3}{k\epsilon_{gas}f(\rm{M})}
\end{equation}
and the mass ratio factor $\mu$, is given by
\begin{equation}
\mu = \frac{M_c}{m}.
\end{equation}
\subsection{Production of bound clusters and the occurrence of mass segregation}
The chief effect of gas dynamical friction is to decelerate objects to sonic speeds. In molecular clouds, turbulence has been observed to be dominated by large modes, whereas sound speeds are rather low and subvirial (see for example \citet{KirkJohnstoneTafalla2007, Kirketal2010}).  We may expect that a cluster will be bound after gas loss, if typical stars -- i.e. those around the peak of the IMF, which carry most of the mass in a cluster -- can lose energy in a time shorter than the gas expulsion time-scale $t_E$ \citep{Gieles2010}.  In our case this would ensure that all higher mass objects would also be bound. \\\\
Thus, in our scenario, with $t_{cross} \sim (G\rho)^{-\frac{1}{2}}$, we get the condition for the formation of a bound cluster as 
\begin{equation}
15 \tau\mu(\rho_{{M_\odot}pc^{-3}})^{-\frac{1}{2}} < t_{E(Myr)}.
\end{equation}
This yields a critical density for the parent gas cloud, such that the cluster emerges bound after gas loss, as 
\begin{equation}
\rho > \rho_{critical} = 225 (\tau {\mu}/t_{E(Myr)})^2\rm{M_\odot}pc^{-3}.
\end{equation}  
For given $\rho$, $\tau$ and $M_c$, we may rewrite the above inequality to give a critical mass $m_{critical}$  such that, all objects with 
\begin{equation}
m > m_{critical} = 15 (\tau M_c/t_{E(Myr)}) (\rho_{{\rm{M_\odot}pc^{-3}}})^{-\frac{1}{2}}
\end{equation}
would be significantly slowed down by gas dynamical friction. 
 \subsubsection{Estimates for the various factors}
 For a pre-cluster, we take $t_{E}$, the length of time for the pre-stellar embedded phase \citep{LeisawitzBashThaddeus1989}, when gas dynamical friction would be operative \citep{Sanchez-SalcedoBrandenburg2001}, as $5\,\rm{Myr}$.  This is less 
than a quarter of the pre-main sequence contraction time for a one solar mass star \citep{Charbonnel1999}. Stellar winds, which can push away the ambient gas, thus reducing gas dynamical friction, are initiated when the stars reach 
the main sequence. Molecular outflows which have the same effect, have dynamical time scales that are much smaller, of the order of $0.1 \rm{Myr}$ only \citep{Frank1999}. \\\\ 
 We may expect a bound cluster to form if cores with masses down to the order of $1\,\rm{M_\odot}$, which can produce stars with masses up to the peak of the IMF \citep{Larsen1982, WeidnerKroupa2006, Weidner2010}, which is $\approx 1/3\,\rm{M_\odot}$ \citep{Chabrier2003}, have $t_{gdf} < 5\,\rm{Myr}$.  This is because they will now have ample time to get decelerated to sub-virial speeds by gas dynamical friction, even as they migrate to the minimum of the gravitational potential. \\\\
We now make an estimate for $\tau$, for star forming cores in clumps, and for clumps in molecular clouds.  We first consider $f(\rm{M})$.  The observed density and velocity structures in molecular clouds are consistent
with supersonic turbulence driven at the scale of the cloud itself \citep{OssenkopfMcLow2002}. In determining the Mach number $\rm{M}$ for a perturber moving through a turbulent medium, the speed of small scale turbulence may be used instead of the sound speed, as the speed with which pressure perturbations are transmited through the medium \citep{SaiyadpourDeissKegel1997}. From  \citet{Ostriker1999} we see that, from $\sim \rm{M}^3/3$ for $\rm{M}<<1$, $f(\rm{M})$ rises sharply to $\sim ln(vt/r_{min})-2$ across $\rm{M} = 1$ and tends to $ln(vt/r_{min})$ for $\rm{M}>>1$. We see that the time scale will be shorter for motion that is supersonic, since $f(\rm{M})$ is $ > 1$ only for supersonic motion \citep{Ostriker1999}. It will be shorter for objects that have comparatively smaller speeds, since a smaller value for $\beta$, implies, both a smaller amount of energy that need be dissipated, and  a smaller separation between the perturber and its wake.  It depends on the density of the parent cloud via the crossing time, and is shorter for denser clouds.  It decreases as the gas fraction in the cloud increases.  Also, it depends directly on the mass of the parent cloud -- the amount of energy that has to be dissipated will be proportional to $M_c$, since the virial speed is proportional to $\sqrt M_c$ -- and inversely on the mass of the perturber, which decides the dissipation rate.  It depends inversely on $f(\rm{M})$, which is related to the speed with which pressure forces can redistribute the density enhancement in the wake.\\\\
We consider now a distribution of speeds for the particles. Those objects which have supersonic speeds with respect to the large scale streaming motion of the gas surrounding it, can be significantly slowed down, by gas dynamical friction.  For subsonic objects, the gas dynamical friction force is depressed due to the fact that pressure forces are able to restore the density distribution about the perturber, sooner \citep{Ostriker1999}.  We see that the chief effect of gas dynamical friction would be to slow supersonic objects down to sonic speeds. Super-virial objects are likely to escape from the system.  So, from here onwards, we put $\beta = 1$ in the expression for $t_{gdf}$. \\\\
 We now consider $k$, the correction factor for non-linearity.  \citet{KimKim2009} investigating the gas dynamical friction force in the non-linear regime, by high resolution hydrodynamic simulations, had measured the strength of the induced gravitational perturbations, by a dimensionless parameter $A = (r_B/r_s) = (Gm/c_s^2r_s)$ ($r_B$ is the Bondi radius). $A$, which corresponds roughly to the ratio of the perturbed density at a distance $r_s$ from the perturber, to the background density, is $>>1$ in the non-linear regime.   Following them, in our case, we determine the correction necessitated by non-linearity, as follows. For a typical  protostellar object of mass $1\rm{M_{\odot}}$,
taking $r_s \sim 100 \rm{AU}$ -the radius of a protostellar nebula, and with $c_s \sim 1\rm{kms^{-1}}$, which is the order of the speed of sound/ random motions in molecular clouds, we get 
\begin{equation}
A = 900 (\rm{kms^{-1}}/c_s)^2 (m/\rm{M_{\odot}})(\rm{AU}/r_s)
\end{equation}
to be $\sim 9$. 
\citet{KimKim2009} obtain $k$ in terms of a non-linearity parameter $\eta = (A/(\rm{M}^2-1))$ as, $k  = (F/F_{Lin})  = (\eta/2)^{-0.45}$ where F is the gas dynamical friction force in the non-linear regime. We convert the limiting values they obtain for $\eta$, into limits on the Mach number $\rm{M}$ for the protostellar object in a star forming gas cloud, as follows.  For $\rm{M} < 1$ and for $\rm{M} > 3.7$  (i.e. $\eta <0.7)$  $(F/F_{Lin}) \sim1$ and for $3.7 > \rm{M} > 2.35$  (i.e. $0.7 < \eta < 2)$, $(F/F_{Lin}) \sim 1.6 -1.0$. For $2.35 > \rm{M} > 1.04$  (i.e. $2 < \eta < 100)$, $(F/F_{Lin}) \sim 1 -1/5$.  We see that for cores which are supersonic, but not greatly so, non-linearity introduces a reduction.  For this range of speeds, we get a typical value for this reduction factor by evaluating it at $\rm{M} = 2$ as  $k = (\eta/2)^{-0.45} \sim 0.83$. For clumps, which are not compact like cores -- with typical masses $\sim 100 \rm{M_{\odot}}$ and sizes $\sim$  few light-year -- we are in the linear regime, and $k=1$.  Thus we have $k \sim 0.83$ and $\epsilon_{gas} \sim 0.9$ for cores in clumps and, $k = 1$ and $\epsilon_{gas} \sim 0.75$ for clumps embedded in molecular clouds.  Here the gas fraction within clouds has been taken as the complement of the mass fraction for clumps in clouds \citep{LadaLada2003, Kauffmannetal2009}.\\\\
For a perturber, which is moving in a turbulent medium, the wake would not be fully developed. Taking $ln(vt/r_{min}) \sim 3$, i.e. for a size of the wake $\sim 20$ times the size of the perturber, $f(\rm{M}) \sim 1$,  and we find that $\tau$ is $\sim 0.05$.  This may be compared with the value $1/4\pi \approx 0.08$ obtained by \citet{Ostriker1999}, as the value for the coefficient relating the gas dynamical friction time scale to, the mass ratio and the typical time scale in the system, while considering the decay of the near-circular orbit of a massive perturber, moving in a massive, spherical cloud with a singular isothermal sphere profile, which result, was itself found consistent with the results obtained by \citet{Sanchez-SalcedoBrandenburg2001}, in three dimensional simulations, of a gravitational perturber moving in a gaseous sphere. 
\section{Comparison with observations}
The formula we have obtained above, for the critical density $\rho _{critical}$, such that, if the cloud gas density $\rho$ is $> \rho_{critical}$, gas dynamical friction would be significant,  for clumps of mass $m$, moving in a cloud of mass $M_c = m/\mu$, may be used to investigate the degree to which gas dynamical friction is significant for cores in clumps and for the clumps themselves, embedded in molecular clouds.\\\\
For a typical clump of mass $500\,\rm{M_\odot}$ -- which, given a SFE $\sim 0.1$ \citep{LadaLada2003}, could be the progenitor of a typical cluster of mass $50\,\rm{M_\odot}$ -- we get a critical density $\sim 1.2\times 10^5\,\rm{cm^{-3}}$.  This is an order of magnitude higher than the density of typical clumps \citep{BerginTafalla2007}, which have densities $\sim 10^3 -10^4\,\rm{cm^{-3}}$.  Also we get the critical mass in typical clouds which have masses $10^3 -10^4\,\rm{M_\odot}$ and densities $50 -500\,\rm{cm^{-3}}$ as, $30 < m_{cr} < 10^3\,\rm{M_\odot}$, the lower value being for less massive clouds with a higher density and vice versa.  Cores which have $m > m_{cr}$, may lose significant amounts of energy, and could congregate (see also \citet{Kirketal2010}) at the potential minima, raising the apparent SFE there.  The compact cluster that forms, can then emerge bound after gas expulsion, if the local SFE, gets raised sufficiently (see discussion before). Thus clusters which were born in sufficiently dense clumps, could emerge bound, and also show mass segregation (since protostars having higher masses and hence, comparatively lower values for $\mu$, would be more strongly affected by gas dynamical friction),  while those born in clumps with comparatively lower density, would form associations.  For a clump of mass $10^3\,\rm{M_\odot}, \rho_{critical}$  is $\sim 5\times 10^5\,\rm{cm^{-3}}$; i.e. an order of magnitude larger, than densities in the upper range of values for typical clumps.\\\\
As a particular example, we now compare the predictions of our theory, with observations of the nearest and well studied Orion Nebula Cluster, which is young enough for the mass segregation observed in it to have been very fast \citep{Reggianietal2011}. It is observed that only stars with masses $\gtrsim 5\,\rm{M_{\odot}}$ are segregated in the ONC \citep{HillenbrandHartmann1998}. \citet{HuffStahler2006} estimate the mass of the parent cloud of the ONC as $6700\,\rm{M_{\odot}}$, within a radius $2.5\,{\rm pc}$, and we get the critical mass (Eqn 9), for the parent cloud of the ONC as $100\,\rm{M_{\odot}}$. From the correlation observed in embedded clusters, between the total mass of the stars in the cluster and the mass of the most massive star in it, the most massive star a clump of mass $100\,\rm{M_{\odot}}$ can produce is  $\approx 3\,\rm{M_{\odot}}$ \citep{WeidnerKroupa2006, Weidner2010}, if we assume a star formation efficiency of $0.3$ \citep{LadaLada2003}. This value is interestingly close to that for the lowest mass observed to be segregated in the ONC. It may be noted that the above mentioned correlation, seen between the total mass of the stars in a cluster and the mass of the most massive star in it, need not conflict with the observations, either of isolated $\rm{O}$ stars or, with numerical simulations which suggest that, clusters form by random sampling from a universal IMF with a fixed stellar upper-mass limit \citep{Lamb2010}. In the first instance, these could be 'run aways', and in the second instance, the correlation would still hold, albeit statistically (see also \citet{Oh2012}).\\\
The density of the parent cloud of the ONC is only $\sim 2\times 10^3{\rm{cm^{-3}}}$, which in our scenario, is less than the critical density ($\lesssim10^6\,\rm{cm^{-3}}$) required for a bound cluster to form, from a cloud of mass $6700\,\rm{M_{\odot}}$ (Eqn 10). Thus, in our scenario, the ONC will evolve into an unbound OB Association and this is consistent with the predictions \citet{HuffStahler2006} make, regarding the fate of the ONC, from observations of the velocity dispersion of the stars in the cluster \citep{JonesWalker1988}.\\\\
From the expression for the critical mass (Eqn 10), we see that, in our scenario, the lowest mass which will be segregated, depends directly on the total mass of the clump, inversely on the square root of the density of the clump and, will be smaller, for smaller speeds of random motions of the gas.  This last dependence, on the level of small scale turbulence, comes from the fact that gas dynamical friction can slow condensations down to the sound speed (here the speed of small scale turbulence), and the lower the speed of the condensations the smaller will be the two body relaxation time (see Eqn 12 below).  The above expectations from our theory, are bourne out by observations \citep{SchmejaKumarFerreira2008, Chavarriaetal2010}, wherein, the most centrally condensed and mass segregated clusters are found in clouds with the lowest Mach number and vice versa.  By making a comparison between the age of the cluster and the migration time scale for the stars, \citet{Chavarriaetal2010} attribute, the mass segregation present in the embedded clusters observed by them, to gas dynamical friction. The densities of these  star forming clumps (which are $ \gtrsim 10^4\rm {cm^{-3}}$), are less than the critical densities (Eqn 9) estimated for their masses (which are $\gtrsim5\times 10^3\,\rm {M_{\odot}}$), and we find that, within them, stars with masses
$\approx 1\,\rm{M_{\odot}}$ do not show any concentration towards the center. We do not expect any of the four clumps observed by \citet{Chavarriaetal2010} to produce bound clusters. However the clump DR21(OH) observed by \citet{Bontempsetal2011}, with a mass of $7000\,\rm{M_{\odot}}$ and a radius of $0.3\,{\rm pc}$ has a critical mass of only $4.2\,\rm{M_{\odot}}$ and, we consider it highly probable that this cloud will eventually produce a bound cluster. 
\section{Discussion}
In our scenario, clusters with large masses, could form at the centers of  sufficiently dense molecular clouds, by the congregation of the more massive clumps in it, due to decelaration by gas dynamical friction.  Observationally some molecular clouds as well as Giant Molecular Clouds, do show, such dense, cluster-bearing concentrations of gas in them \citep{WilkingLada1983, Maiz2001}, and \citet{WilkingLada1983} comment on their probable significance in the formation of bound stellar clusters.  Also, gas that harbors clusters, is seen to be more dense \citep{Kauffmannetal2010} and more highly clumped or substructured than gas that does not bear a cluster \citep{LadaLada2003}.  In the present scenario, the cluster will bear the imprint of the merged clumps and cores, as substructure in its stellar distribution, till it completely relaxes. This conforms with what is observationally seen, since very young, ($< 1\rm{Myr}$) star clusters, seem to show significant levels of substructure \citep{LadaLada2003} and, observations show that cores and young clusters have substructure and are cool (sub-virial) (see for example \citep{SchmejaKumarFerreira2008, KirkJohnstoneTafalla2007}).  This is consistent also, with the observed difference, between the fractal dimensions of young clusters with internal substructure (that has still not been erased by dynamical relaxation) and, that of molecular gas, and we note that, a more clustered distribution of the densest star forming gas, within the parent molecular clouds -- as expected in our scenario -- has already been suggested as a probable reason for the above difference in fractal dimensions \citep{AlfaroSanchez2011}.  Mergers, need not erase any mass segregation that was originally present within the clumps, since dynamical simulations have shown that, in the merging of clusters, which are themselves mass segregated, the mass segregation will be retained, rather than get  erased in the merger \citep{McMillan2008}.   In this context, it may be interesting to note that the comparatively less dense Taurus complex, does not show any central concentration or mass segregation.\\\\
The critical densities above which we expect the formation of bound clusters / mass segregation, are consistent with the inter-clump medium densities, for which \citet{GortiBhatt1995} notice mass segregation in their numerical simulations studying the effect of gas dynamical friction on prestellar clumps and, \citet{SaiyadpourDeissKegel1997} obtain bound clusters in their analysis.  The larger densities, which in our scenario are a prerequisite for the formation of bound clusters, are also consistent with the general conclusion that the production of a bound cluster must require special physical conditions, to account for their rarity; only about $4 -7\%$ of all embedded clusters seem to give rise to bound clusters \citep{LadaLada2003}. Also \citet{Kauffmannetal2010} report that, between clouds which have the same radii, those which harbor clusters are more massive than those which do not show a concentration of stars.  They also report that while clouds do seem to follow a general mass -size relation of the form $M_c(r) = 400 \rm{M_{\odot}}{r_{pc}}^{1.7}$, there are outliers whose masses (and hence densities) are an order of magnitude larger.  In our scenario, such denser clouds could give rise to bound clusters. \citet{GBK2010} in a study, on the fraction of clusters that escape infant mortality find that, in the galaxies they studied, this fraction is strongly proportional to the Star Formation Rate (SFR) density of the parent galaxy.  Since the SFR density is related to the surface gas density by the Kennicutt-Schmidt law, they conclude that the survival fraction of clusters is strongly dependent on the surface gas density.  Since we expect clouds, and hence clumps, in regions of larger surface gas densities to be denser, this is consistent with the expectation from our scenario. Our model is consistent also with the observed correlation, that the two categories of massive clusters -- bound or 'starburst' and unbound or 'leaky' -- noticed by \citet{Pfalzner2009} obey; that bound clusters are found in regions of high density viz. the Galactic Center or in the spiral arms, and unbound ones in the lower density areas \citep{Pfalzner2011}. She even suggests, as a hint for the origin of the two categories, this obvious difference in their locations. It is also interesting to note that while, among these clusters, the young, bound ones have stellar densities $ > 10^5\,\rm{cm^{-3}}$, the initial stellar densities for the young, unbound ones are $ < 5\times 10^3\,{\rm{cm^{-3}}}$. The high initial densities quoted for the bound 'starburst' clusters -- $\sim 10^6 -10^7\,\rm{cm^{-3}}$ -- \citep{Pfalzner2009, PfalznerEckart2009} are consistent with the densities expected for the compact clusters, that in our scenario, should form in clouds with densities greater than the critical densities we obtain (Eqn 9) (see also \citet{Maiz2001}).\\\\
The  gas dynamical friction time scale is inversely proportional to the mass of the body which is being decelerated by it. Hence it is possible that, relative segregation, between the stars / cores / clumps of various masses may occur and, some signature of this might survive, even after relaxation, in the clusters evolving out of dense gas clouds. We note that, the brown dwarfs associated with a cluster are distributed in a region much larger than what is occupied by the cluster itself \citep{Luhman2006, KumarSchmeja2007, Kouwenhoven2007}, and clusters do seem to have associated with them a corona of low mass stars (\citep{Sharmaetal2006, Luhman2006, KumarSchmeja2007, Kouwenhoven2007}; see also \citet{GortiBhatt1996}).  Such a 'core-halo' structure has been noticed for some Massive Young Clusters too \citep{Maiz2001}. \\\
It is seen from numerical simulations that, clusters that are dynamically cool at birth can remain bound, down to SFE's $\sim 0.05$, if they lose the gas while they are still cool \citep{GOODWIN1997, Goodwin2009, SMITHETAL2011} and that, clusters that are cool and/or have substructure, can mass segregate in a very short time scale \citep{Allisonetal2009, Allison2010, MASCHBERGER2011, OLCZAKSPURZEMHENNING2011}.  However, the study by \citet{PelupessyPortegiesZwart2012}, which extended these studies by including the potential due to the gas, found that, only a remnant with an IMF that was possibly skewed, rather than a mass segregated cluster, is left after gas expulsion. That bound clusters may be obtained, by starting with 'cool' initial conditions, and losing the gas before the stars acquire too high speeds due to collapse, may be expected from the fact that, between being born with a very low velocity dispersion, and acquiring virial speeds by collapsing within the gravitational potential of the gas cloud -- within about a dynamical time, which is $\sim$ half a million year, for a density $\sim 10^4\,\rm{cm^{-3}}$ --  \citep{LadaMargulisDearborn1984}), the stars can,  for a short time, satisfy the condition for binding -- that the total energy of the stars, taken alone, be negative. However, this would imply that the survivability of clusters would be proportional to the dynamical time, which is inversely proportional to the square root of the density. This would produce an anti-correlation between the cluster survivability and the gas density, which is counter to observations, as explained above \citep{GBK2010}.\\\
The scenarios mentioned above, which produce fast mass segregation, may be understood by noticing the following.  The formula for the two body dynamical relaxation time for a stellar cluster, whose stars are moving with virial speeds \citep{Binney2008}, has to be modified when we consider a cluster that is embedded in gas. With $\epsilon$ equal to the ratio of the total mass in stars to the total mass of stars plus gas we get
\begin{equation}
t_{Erelax} =
\frac{\alpha ^2N}{8{\rm{ln}} \alpha N}\,t_{cross}
\end{equation} 
where $\alpha = \beta ^2 / \epsilon$.  $\epsilon$ is $ < 1$ for an embedded cluster and, is equal to $1$ for a cluster free of gas.  The relaxation time would be smaller for -- a) larger SFE /gas free clusters  b) sub-virial speeds for the stars i.e. $\beta < 1$, and also c) within sub-structure, since in this case, the number of stars as well as the crossing time would be much smaller compared to those for the cluster as a whole.  These conclusions are in concurrence with the results of the simulations mentioned above, wherein, the signature of dynamical relaxation viz. mass segregation, is reported to occur fast, in simulations which start with suitable (as per our analysis above) initial conditions.  Our model, explores the possibility of gas dynamical friction being the probable cause of the occurrence of 'suitable' initial conditions, and also delineates the situations in which such initial conditions might arise.\\\
In our scenario, the difficulty posed by the large dynamical relaxation times in embedded clusters, is avoided by the fact that, gas dynamical friction is capable of poducing mass segregation explicitly -- due to its mass dependence, as well as implicitly too -- by reducing the dynamical relaxation time, through its ability to decelerate condensations to sonic speeds, which in star forming clouds, are subvirial (i.e. by making $\beta$ go $ < 1 $) (Eqn 12).  The efficacy of gas dynamical friction may be enhanced, under the following conditions. In turbulent molecular clouds, the decay time for small scale turbulence, will be smaller than that for large scale motions, if $m_t < 3$, where $E_t(k_t) \propto k_t^{-m_t}$ is the energy density of turbulent motions per unit wave number range, $k_t$ being the wave number.  For example, for a Kolmogorov spectrum of turbulence, $m_t = 5/3$. Hence the sound speed, which is here taken as the typical speed of small scale turbulence in the cloud, can decrease over time and go over to the actual speed of sound for quiescent gas, in situations where the turbulence is neither forced nor fed, and clumps and cores, which are slowed down to sonic speeds by gas dynamical friction, can be quite sub-virial. This will reduce the relaxation time. The dying down of turbulence can cause the friction to increase also, by allowing longer wakes to develop. Also, for a cloud with a power law density gradient $\propto r^{-m_d}$  ($m_d<3$), between a small inner radius within which the density is a constant, and a large outer radius beyond which the density may be considered negligible, we get $M_c(r)<\rho(r)>^{-1/2}\propto r^{3-\frac{m_d}{2}}$, $<\rho(r)>$ being the mean density within a radius $r$. Thus the gas dynamical friction time scale is likely to be smaller than the global mean value towards the centres of real clumps and clouds, as the product $M_c {\rho}^{-1/2}$, which occurs within the equation for $t_{gdf}$ (Eqn. 4), decreases with $r$.\\\\
Though gas dynamical friction can cause decay of binary orbits \citep{GortiBhatt1996{a}} and, binaries and stellar clusters can have significant effects on each other \citep{Kroupa1995, Kroupa1999, Delgado-Donateetal2003, Parkeretal2011{a}}, we do not consider the possible role, that either binaries or tidal fields (see for example \citet{Ballesteros-Paredes2009}), could play in the formation of bound clusters, as they lie beyond the scope of this paper.  The model avoids the need, to take recourse to density dependent IMF's or SFE's \citep{Kruijssen}, in accounting for various  aspects of stellar clusters.   A comparison with such models will be made in a future work.\\\
The model offers the following interesting possibility for the formation of Globular Clusters -- by the contraction and coalescence, in a similar scenario, of dense cores, within clumps, within clouds, within very massive, very dense molecular clouds, that perhaps formed via turbulent fragmentation during galaxy formation (see also \citet{Maiz2001}).
\section{Summary}
Gas dynamical friction can decelerate masses and also explicitly as well as implicitly, promote mass segregation.   
Comparing the gas dynamical friction time-scale for sub-structures within star forming clouds, with the time available before gas expulsion takes place from an embedded stellar cluster, we obtain a boundary value for the density of a star forming clump of given mass, such that, stellar clusters born in clumps which have densities higher than this, could emerge bound after gas loss, due to the locally elevated SFE.  For a clump of given mass and density, we find a critical mass such that, subcondensations with larger masses than this could suffer significant segregation within the clump.  We compare our results with observations and also other work in the field, and also discuss the implications of our scenario for the formation of bound clusters, vis-a-vis observations.  
\section*{Acknowledgements}
The author thanks Chanda Jog and Rajaram Nityananda for many useful discussions. She thanks the University Grants Commission for a Visiting Fellowship to the Indian Institute of Science (Bengaluru) and also, the Inter University Center for Astronomy \& Astrophysics (Pune), for an Associateship.
\label{lastpage}

\end{document}